\documentclass{article}

\usepackage{arxiv}

\usepackage{amsmath,amsfonts}
\usepackage{algorithmic}
\usepackage{algorithm}
\usepackage{array}
\usepackage[caption=false,font=normalsize,labelfont=sf,textfont=sf]{subfig}
\usepackage{textcomp}
\usepackage{stfloats}
\usepackage{url}
\usepackage{verbatim}
\usepackage{graphicx}
\usepackage{cite}
\usepackage{bm}
\usepackage{mathtools}

\newcommand{\bx}{\bm{x}}
\newcommand{\bk}{\bm{k}}

\newcommand{\hr}{SST_{\text{HR}}}
\newcommand{\lr}{SST_{\text{LR}}}
\newcommand{\aen}{SST_{\text{AE}}}
\newcommand{\gan}{SST_{\text{GAN}}}

\begin{document}

\title{Super-resolution of satellite-derived SST data\\ via Generative Adversarial Networks}

\author{Claudia Fanelli$^1$, Tiany Li$^2$, Luca Biferale$^2$, Bruno Buongiorno Nardelli$^1$, Daniele Ciani$^3$, Andrea Pisano$^3$ \& Michele Buzzicotti$^2$}

\author{
Claudia Fanelli \\
CNR-ISMAR \\
Calata Porta di Massa, 80133, Naples, Italy\\
 \And
Tiany Li\\
University of Rome Tor Vergata\\
Via della Ricerca Scientifica 1, 00133 Rome, Italy\\
\And
Luca Biferale\\
University of Rome Tor Vergata\\
Via della Ricerca Scientifica 1, 00133 Rome, Italy\\
\And
Bruno Buongiorno Nardelli \\
CNR-ISMAR \\
Calata Porta di Massa, 80133, Naples, Italy\\
\And
Daniele Ciani \\
CNR-ISMAR \\
Via del Fosso del Cavaliere 100, 00133, Rome, Italy\\
\And
Andrea Pisano\\
CNR-ISMAR \\
Via del Fosso del Cavaliere 100, 00133, Rome, Italy\\
\And
Michele Buzzicotti\\
University of Rome Tor Vergata\\
Via della Ricerca Scientifica 1, 00133 Rome, Italy\\
}

{Fanelli \MakeLowercase{\textit{et al.}}: Super-resolution of satellite-derived SST data via Generative Adversarial Networks}

\maketitle

\begin{abstract}
In this work, we address the super-resolution problem of satellite-derived sea surface temperature (SST) using deep generative models. Although standard gap-filling techniques are effective in producing spatially complete datasets, they inherently smooth out fine-scale features that may be critical for a better understanding of the ocean dynamics. We investigate the use of deep learning models as Autoencoders (AEs) and generative models as Conditional-Generative Adversarial Networks (C-GANs), to reconstruct small-scale structures lost during interpolation. Our supervised -model free- training is based on SST observations of the Mediterranean Sea, with a focus on learning the conditional distribution of high-resolution fields given their low-resolution counterparts. We apply a tiling and merging strategy to deal with limited observational coverage and to ensure spatial continuity. Quantitative evaluations based on mean squared error metrics, spectral analysis, and gradient statistics show that while the AE reduces reconstruction error, it fails to recover high-frequency variability. In contrast, the C-GAN effectively restores the statistical properties of the true SST field at the cost of increasing the pointwise discrepancy with the ground truth observation. Our results highlight the potential of deep generative models to enhance the physical and statistical realism of gap-filled satellite data in oceanographic applications.
\end{abstract}

\section{Introduction}\label{sec:intro}
Covering more than 70\% of Earth’s surface, the world’s oceans play a fundamental role to regulate climate variability and providing essential services to human life. The unprecedented anthropogenic pressure over the environment during the last decades lead to several consequences on the ocean’s health, impacting on its physical and biogeochemical properties. One of the most severe changes consists in the continuous increasing of sea temperature, with a global warming rate estimated between 0.13  and 0.22 $\pm$ 0.01 $^{\circ}$C per decade \cite{von2024state,bbn2025dynamical}, influencing marine habitats, climate variability, global circulation and ocean-atmosphere interaction \cite{bbn2025dynamical,mackenzie2007long,jha2014sst,dong2018asymmetric,woollings2010storm,bentamy2017review}. For this reason, constant and global observation of sea temperature is essential to assess the state of the ocean and the repercussions on its health. At the regional level, the Mediterranean Sea is especially prone to rapid warming, experiencing a warming rate equal to 0.41 $\pm$ 0.01 $^{\circ}$C per decade \cite{pisano2020new}, making it a perfect hot spot for monitoring global changes \cite{giorgi2006climate}.

\begin{figure*}[!t]
\includegraphics[width=\textwidth]{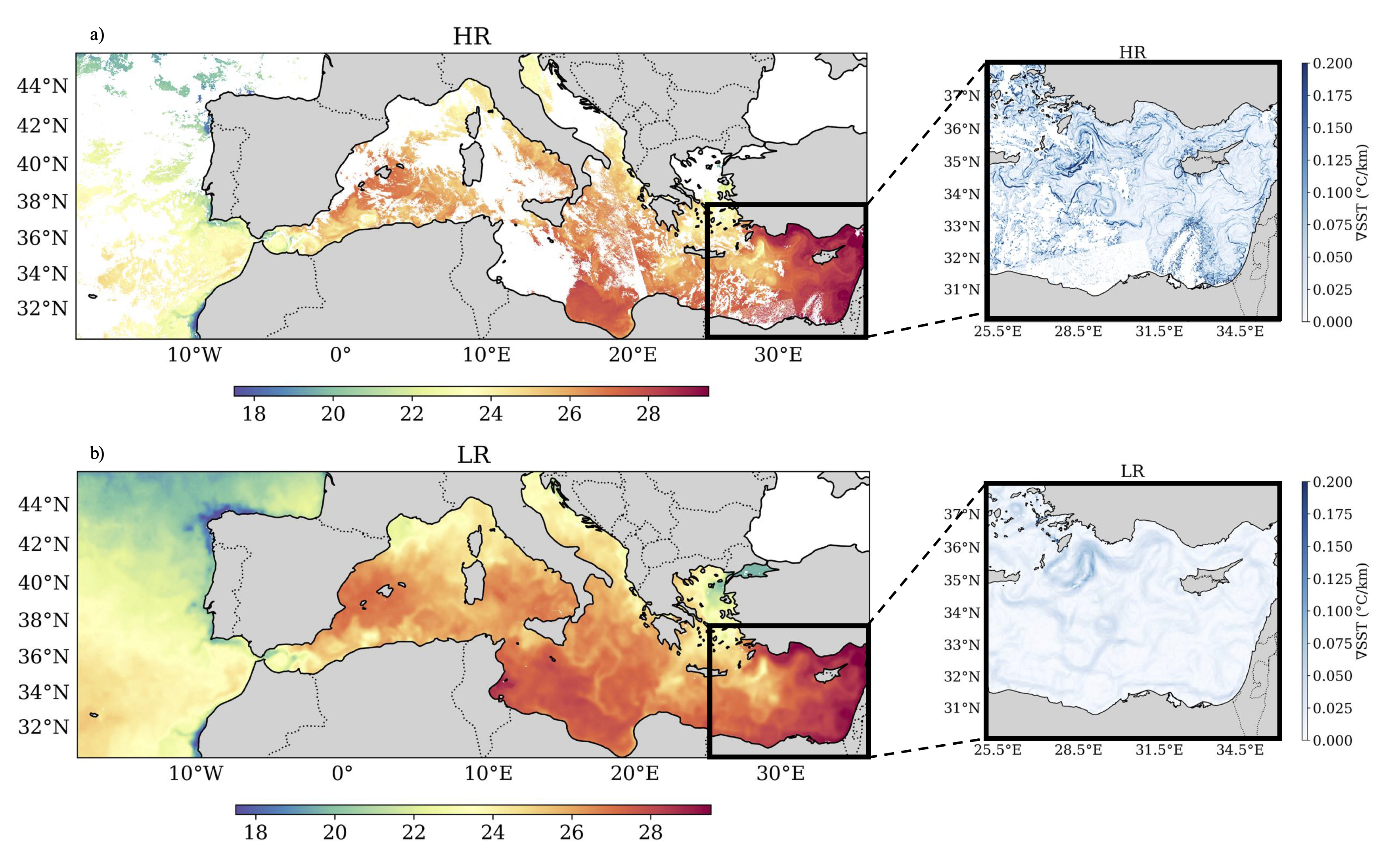}
\caption{SST fields over the Mediterranean Sea described by (a) L3S data obtained from the supercollection and processing of Sentinel 3A and 3B granules and (b) the Near Real Time gap-free L4 product provided by the Copernicus Marine Service at 1/16° spatial resolution and remapped over a 1/100° regular grid for the day 10/09/2021 (julian day = 253). The regional zooms show the SST gradients over the Levantine Sea to highlight the impact of the interpolation technique.\label{fig:HRvsLR}}
\end{figure*}

High-resolution sea surface temperature (SST) data obtained from remote sensing systems have revolutionized our ability to monitor and understand many dynamical and biological processes at both large and small scales in the ocean and at the air-sea interface. While the global coverage of satellite measurements is crucial to observe the large scale bio-physical dynamics, high-resolution data allow to capture small scale features essential to a wide variety of applications. In fact, at mesoscale and submesoscale ranges, it is well known that the dynamical information embedded in SST data can be leveraged to improve the spatial and temporal resolution of the geostrophic current fields, e.g. \cite{bowen2002extracting,ciani2020improving,isern2021high}, and to evaluate the bio-physical response to ocean warming in important ecological regions such as coastal areas, e.g. \cite{weidberg2023assessing, fanelli2024using}. Capturing small spatial variations of SST also plays a significant role when dealing with meteorological models, given the established response of marine atmospheric boundary layer to small SST perturbations, e.g. \cite{koravcin1990numerical,oneill2005high}, and the proven sensitivity of extreme weather events modeling to small variations of SST, e.g. \cite{hirons2018impact, avolio2021multiple}. This research area is also driving the design of future space-born Earth observations systems \cite{weidberg2021global,liberti2023multi}. Having such a fine and comprehensive view of one of the essential ocean and climate variables (EOV/ECVs) allows us to capture multiscale thermal variations and evaluate the state of the ocean. However, the wide variety of available products that describe the SST over the global and regional oceans at different spatial and temporal resolutions are all affected by the limitations of the instruments used to measure the temperature, which often force the scientific community to choose between high-resolution data (i.e., infrared-based measurements, affected by cloud cover) or lower resolution, almost all-weather vision (i.e., microwave sensors, mainly affected by precipitations). Reaching scales ranging from kilometers to sub-kilometers, the former are usually preferred, leaving the data producers/users to deal with data voids caused by cloud coverage (Fig.~\ref{fig:HRvsLR}a). However, even the most complex statistical techniques used to interpolate geophysical data at unsampled locations come with a series of weaknesses, resulting in the production of relatively smooth fields (Fig.~\ref{fig:HRvsLR}b). Due to the spatial and temporal averaging of these methods, which tends to flat high frequency variability, the smoothing effect can produce the loss or the under-representation of small scale features. The regional zoom of the SST gradients over the Levantine Sea illustrated in Fig.~\ref{fig:HRvsLR} shows the impact of the gap-filling procedure, which averages out the small-scale features present in the original gappy satellite observations. 

In the last years, several attempts have been made to improve the accuracy or fill the gaps of remotely-sensed sea surface temperature data, exploiting the potential of deep learning algorithms. Machine learning techniques are in fact particularly suitable to identify non-linear patterns and spatiotemporal correlations within large datasets, widely offered by the long-term and global observations made available by satellite measurements. Several works took advantage of the physical correlation with other oceanographic variables, such as altimetry observations \cite{buongiorno2022super, barth2022dincae, archambault2024learning, ciani2025estimating} and chlorophyll distribution \cite{hayes2005glacial, sammartino2020artificial, cutolo2024cloinet}, to help networks reconstruct or improve the resolution of SST data. Moreover, various authors have explored the possibility of supporting numerical predictions through the combination of models with deep neural networks, e.g. \cite{garcia2007prediction, meng2023physical, haghbin2021applications}. Only recently, a few authors have focused on improving the spatial resolution of SST satellite observation using machine learning-based approaches \cite{manucharyan2021deep, lloyd2021optically, martin2023synthesizing, zou2023super, fanelli2024deep}.
The emergence of data-driven super-resolution techniques, such as generative adversarial networks (GANs)~\cite{goodfellow2014generative}, and diffusion models~\cite{ho2020denoising}, has shown significant promise in solving this task beyond traditional interpolation methods. These data-driven approaches are particularly attractive in geoscience~\cite{lockwood2024generative, martin2024deep}, where we do not possess the exact model, and it is mandatory to learn from empirical ground-truth observations. By learning from high-resolution input, generative models can reconstruct missing data and infer subgrid-scale variability, offering a powerful framework for both super-resolution and gap-filling tasks~\cite{buzzicotti2023data}. Their ability to capture the underlying statistical structure of complex physical systems, \cite{li2024synthetic}, opens new avenues to improve the quality and realism (in terms of their statistical fluctuations) of Earth system datasets.

In this work, our goal is to explore the potential of deep conditional generative models to super-resolve low-resolution (LR) gap-free maps and generate high-resolution (HR) gap-free SST data. 
To supervise the training of conditional generative models in reconstructing small-scale statistics of the LR input maps, we use data from the regions where HR observations are available. Preparing a sufficiently large and high-quality dataset for training, validating, and testing a deep-generative model to achieve such a goal is particularly challenging.  Unlike many previous case studies, where datasets have been specifically generated using model-based numerical simulations,  our work relies on satellite data, which are inherently limited, but model-free. First, the amount of data available is limited by the duration of the satellite mission, which means that the number of HR gap-filled tiles cannot be arbitrarily increased to meet training needs.  Second, the ground truth itself is subject to real observational uncertainties, as satellite measurements are affected by atmospheric conditions and sensor limitations. These factors make it very difficult to curate a robust and comprehensive dataset for training deep generative models.
To mitigate these limitations, the first strategy is to optimize the tile size used as input/conditioning and output of the generative model implemented in our super-resolution tasks. Moreover, the power spectral density (PSD) profile of the interpolated SST compared to the ground-truth data presented in \cite{fanelli2024deep} shows the same behaviour for wavenumbers smaller than 1 per degree, which means for scales larger than 100 km. As a compromise between having a sufficient number of HR, gap-free training samples large enough to preserve the mesoscale structures necessary for meaningful conditioning, we choose tiles of $1^\circ \times 1^\circ$ degree.
The second strategy is to apply super-resolution not directly to the raw satellite SST dataset, but rather to its anomaly field. Specifically, we preprocess the SST dataset by subtracting the local mean temperature of each tile and focus only on the anomalies. This allows us to aggregate data from different regions into a larger, spatially independent training dataset, improving both the data volume and the generalizability of the generative model.
By training on these spatially diverse anomaly tiles, the generative model can learn to recover fine-scale details in a way that is independent of regional biases in absolute SST values. After super-resolving the anomalies, the original mean can be added back to each tile to obtain the true local SST HR map.


\section{Methods}
\label{sec:methods}


\subsection{Training and test datasets}
\label{subsec:datasets}

The SST fields fed to the networks are derived from the Mediterranean Sea High Resolution (HR) and Ultra High Resolution (UHR) Sea Surface Temperature Analysis freely distributed within the Copernicus Marine Service at near real time (\url{https://doi.org/10.48670/moi-00172}). These data provide the foundation SST, nearly free of diurnal variability, at daily temporal resolution from 2008 to one day before real time, and are based on nighttime images collected by infrared sensors from different satellite platforms. After a series of operations, which includes quality control, image merging, cloudy pixel removal, and bias correction, two different spatial resolutions are available for the merged multisensor (Level 3, L3) and gap-free (Level 4, L4) datasets over the Mediterranean Sea: a coarser product at 1/16$^\circ$ nominal resolution and a finer one over a 1/100$^\circ$ regular latitude-longitude grid.
The 1/100$^\circ$ L4 SST fields are obtained with a gap-filling algorithm based on a two-step  optimal interpolation (OI) scheme \cite{nardelli2013high} and applied to the observational L3 data. The first step consists in carrying out a HR OI processing to fill in the large data gaps (either due to cloudy areas and insufficient coverage by infrared sensors), which is thus run at 1/16$^\circ$. This first SST field is successively upsized at 1/100$^\circ$ to be used as the background to the second UHR OI (that imposes much smaller spatial and temporal decorrelation scales). While this second step is able to keep kilometric scales features when effectively observed in available images, areas originally covered by clouds only resolve the signals effectively kept in the initial 1/16$^\circ$ fields. As such, our aim here is to improve the UHR OI background field by applying different super-resolution algorithms. This will eventually improve the OI reconstruction of the UHR field in the entire domain. The target dataset is then specifically built by applying the processing chain to data acquired by the Sea and Land Surface Temperature Radiometer (SLSTR) aboard the Sentinel-3A and 3B satellites in order to obtain merged and high accurate images of SST at 1/100$^\circ$ spatial resolution. The final training dataset is composed of pairs of tiles of dimensions $1^\circ \times 1^\circ$ (with overlaps of $0.5^\circ$) extracted from the input/target images of SST over the Mediterranean Sea during the year 2020 following the steps indicated hereafter. First, given the fact that the target data still have many gaps due to the inability of the infrared sensors to see through clouds, we selected only tiles with at least 95\% valid pixels. Successively, the spatial mean of each tile is subtracted from SST values to convert them into anomalies and eliminate seasonal variations, and finally each tile is normalized to a range of -1 to 1 using a standard global min-max scaling technique.

The final training dataset is composed by 94110 pairs of tiles, while the test dataset is built following the same procedure, excepting the exclusion of data with gaps, using 20 days from the year 2021 which produce 92880 tiles (where 4728 are the tiles with at least 95$\%$ of valid pixels and roughly 400 the tiles with no gaps).


\subsection{Super resolution for geophysical data}
\label{subsec:sr}

One must be careful when addressing a super-resolution task, as it is an inherently ill-posed problem. The number of degrees of freedom to be reconstructed far exceeds those observed at large scales. Consequently, it is not possible to uniquely determine the solution to the problem. This implies that for any given LR map, there exist infinitely many physically consistent HR reconstructions that satisfy the same LR constraints.
As a result, neither data-driven nor equation-based approaches can achieve zero MSE in small-scale reconstruction, or, in geophysical terms, can reach the exact forecast of the regional fluctuations based on the mesoscale dynamics. This highlights the intrinsically probabilistic nature of super-resolution, where the relevant question is not how to deterministically reconstruct a unique HR field but rather: What is the probability that, given a specific large-scale configuration, certain high-frequency fluctuations will be observed?
Specifically, what we can aim to model is the conditional distribution of the data, $p_{data}(\bm{x}|\hat{\bm{x}})$, where $\hat{\bm{x}}$ represents the LR map and $\bm{x}$ the corresponding HR data.
To answer this question, we aim to learn the conditional distribution from the available HR satellite observations, which, despite their sparsity when accumulated over time, can provide sufficient statistical information to approximate the HR probability distribution over the entire observed region.
Since $p_{data}(\bm{x}|\hat{\bm{x}})$ is entirely unknown and defined in the high-dimensional space determined by the support of the HR data, it is infeasible to approximate or fit it using physics-informed methods.\\
\noindent 
Recently, data-driven generative models have been developed relying on the assumption that each data point of a given training set is drawn from the same underlying probability distribution, whose functional form is generally unknown. To be more precise, let $\left \{ \bm{x}_1, \bm{x}_2, ..., \bm{x}_N \right \}$ denote a dataset of size $N$, the previous statement says that we can see all $\bm{x}_i$ as an independent and identically distributed (i.i.d.) sample extracted from $p_{data}(\bm{x})$. 
Thus, the goal of generative modeling is to estimate the unknown probability distribution underlying a given data set so that we can generate new data samples at will. To this end, we build a statistical model $p_{\theta}(\bm{x})$, where $\theta$ denotes the model parameter, and we seek to find the optimal parameter $\theta^*$ such that $p_{\theta^*}(\bm{x}) \approx p_{data}(\bm{x})$.
Before extending this approach to the specific problem of super-resolution and conditional generation in general, we first examine how generative models build and optimize such a probability distribution.
The optimization is done by minimising some metric $D(\cdot || \cdot)$ that measure the difference between $p_{\theta}(\bm{x})$ and $p_{data}(\bm{x})$,
\begin{equation}
\label{eq:theta*}
\theta^* \coloneq \arg \min_\theta D(p_{\text{data}}(\bx) \| p_{\theta}(\bx)).  
\end{equation}
A common choice for $ D(\cdot || \cdot) $ is the class of $ f$-divergences \cite{song2021train}, defined as:
$$
D_f(p_{\text{data}}(\bx) \| p_{\theta}(\bx)) = \int p_{\theta}(\bx) f \left( \frac{p_{\text{data}}(\bx)}{p_{\theta}(\bx)} \right) d\bx.
$$
For any convex function $ f: [0,\infty) \to \mathbb{R} $ with $ f(0) = 1 $, this divergence is always non-negative and equals zero if and only if $ p_{\text{data}}(\bx) $ and $ p_{\theta}(\bx) $ are identical. A notable example, widely used in machine learning, is the Kullback-Leibler (KL) divergence obtained by selecting $ f(x) = x \log(x) $ \cite{song2021train}:
\begin{equation}
\label{eq:DKL}    
D_{KL}(p_{data}(\bx)||p_{\theta}(\bx)) = \int p_{data}(\bx) \log \left ( \frac{p_{data}(\bx)}{p_\theta(\bx)} \right ) d\bx. \end{equation}
It is possible to notice that by replacing~\eqref{eq:DKL} into~\eqref{eq:theta*}, and by approximating the expectation in~\eqref{eq:DKL} with the empirical mean over the training samples, we can effectively approximate $p_{data}(\bx)$ in a supervised way solely using the training dataset,
\begin{align}
\nonumber
D_{KL}(p_{data}(\bx)||p_{\theta}(\bx)) &\approx \frac{1}{N}\sum_{i=1}^N \log \left ( \frac{p_{data}(\bx_i)}
{p_\theta(\bx_i)} \right )  \\
\nonumber
& = \frac{1}{N}\sum_{i=1}^N \log p_{data}(\bx_i) \\
\nonumber
& - \frac{1}{N}\sum_{i=1}^N \log p_{\theta}(\bx_i) \\
& = -\frac{1}{N}\sum_{i=1}^N \log p_{\theta}(\bx_i)  + const,
\label{eq:DKLapprox}
\end{align}
where in the third step the term $\frac{1}{N}\sum_{i=1}^N \log p_{data}(\bx_i)$ is set to a constant, since it does not depend on the model parameter and does not enter in the model training.
Eq.~\eqref{eq:DKLapprox} states that training a generative model is equivalent to maximizing the log likelihood of the model distribution over all points in the dataset,
\begin{equation}
\label{eq:theta*2}
\theta^* \approx \arg \max_\theta \frac{1}{N}\sum_{i=1}^N \log p_{\theta}(\bx_i).
\end{equation}
A significant challenge in applying generative models to high-dimensional data, such as in our case of turbulent sea surface dynamics, is to ensure that the normalization constraint is satisfied by the modeled probability, i.e., that the total probability over all possible data configurations sums to one.
In the literature there are at least three main approaches to overcome this problem: (i) explicitly normalized models, such as VAEs or Normalizing Flows; (ii) adversarial training, and (iii) score-based models.
In this work, we focus on GANs for the following reasons. First, despite challenges such as limited diversity in generated samples and unstable training \cite{ buzzicotti2023data,metz2016unrolled}, GANs have demonstrated high accuracy in statistically reconstructing turbulent dynamics \cite{buzzicotti2021reconstruction, li2023multi}, with superior quality compared to VAEs. Second, GANs offer a well-established framework for implementing conditioning, making them a suitable choice given the additional challenge of dataset limitations in this context. Third, conditional GANs can be designed as a deterministic generator that maps LR inputs to HR outputs while optimizing the statistical properties of reconstructed small-scale fluctuations. This deterministic mapping ensures smooth transitions when merging super-resolved tiles, preventing artifacts or discontinuities in the final reconstructed field. In contrast, stochastic sampling from score-based models could introduce inconsistencies across adjacent regions, making them less suitable for our application in their standard formulation. We leave their extension as a next step.
In the following subsections, we detail the modifications made to the standard GAN training framework to enforce conditioning and ensure deterministic HR reconstruction.

\subsection{Conditional Generative Adversarial Network}
\label{subsec:gan}

In this work, we modified the standard generative process~\cite{goodfellow2014generative} to incorporate conditioning. The generator network is transformed into an auto-encoding structure so that, instead of mapping the output data from a random input, it takes the LR map as input and reconstructs the corresponding HR data using the decoder.
Specifically, we implemented the generator as a residual network (ResNet), adding its output to the low-resolution input. This is possible because, in our setup, the LR and HR data share the same $100 \times 100$, ($1/100^\circ$ resolution), spatial grid. This architecture is known to be more stable during training and can be seen as an effective small-scale correction added to the LR input~\cite{he2016identity}. This reformulation defines the generator as a deterministic mapping:
$$
\bm{f_\theta}: \mathbb{R}^d \rightarrow \mathbb{R}^d, \quad \bx_\theta = \hat{\bx} + \bm{f_\theta}(\hat{\bx}),
$$
where $ \hat{\bx} $ represents the LR input and $ \bx_\theta $ is the super-resolved output.\\
\noindent
The generator loss function is also modified to a multi-objective loss comprising two terms; (i) the adversarial loss, ensuring the statistical consistency of the generated samples. (ii) A Mean Squared Error (MSE) loss, which enforces reconstruction fidelity by penalizing deviations from the ground-truth HR data.
Thus, the optimal parameters for both the discriminator ($ \phi^* $) and the generator ($ \theta^* $) in the conditional adversarial training are obtained as:

\begin{equation} \label{eq:discr_cond}
    \phi^* \coloneq \max_\phi \left [ \mathbb{E}_{p_{\text{data}}(\bx)} \log \bm{d_\phi}(\bx) + \mathbb{E}_{p_{\text{data}}(\hat{\bx})} \log (1-\bm{d_\phi}(\bx_\theta))) \right ],
\end{equation}
\begin{equation}\label{eq:gener_cond}
    \theta^* \coloneq \min_\theta \left [ (1-\lambda) \, \mathbb{E}_{p_{\text{data}}(\hat{\bx})} \log (1-\bm{d_\phi}(\bx_\theta))) + \lambda \, \mathbb{E}_{p_{\text{data}}(\hat{\bx})} \left \| \bm{x} - \bx_\theta \right \|_2 \right ].
\end{equation}

\begin{figure*}[h!]
\includegraphics[width=1.\textwidth]{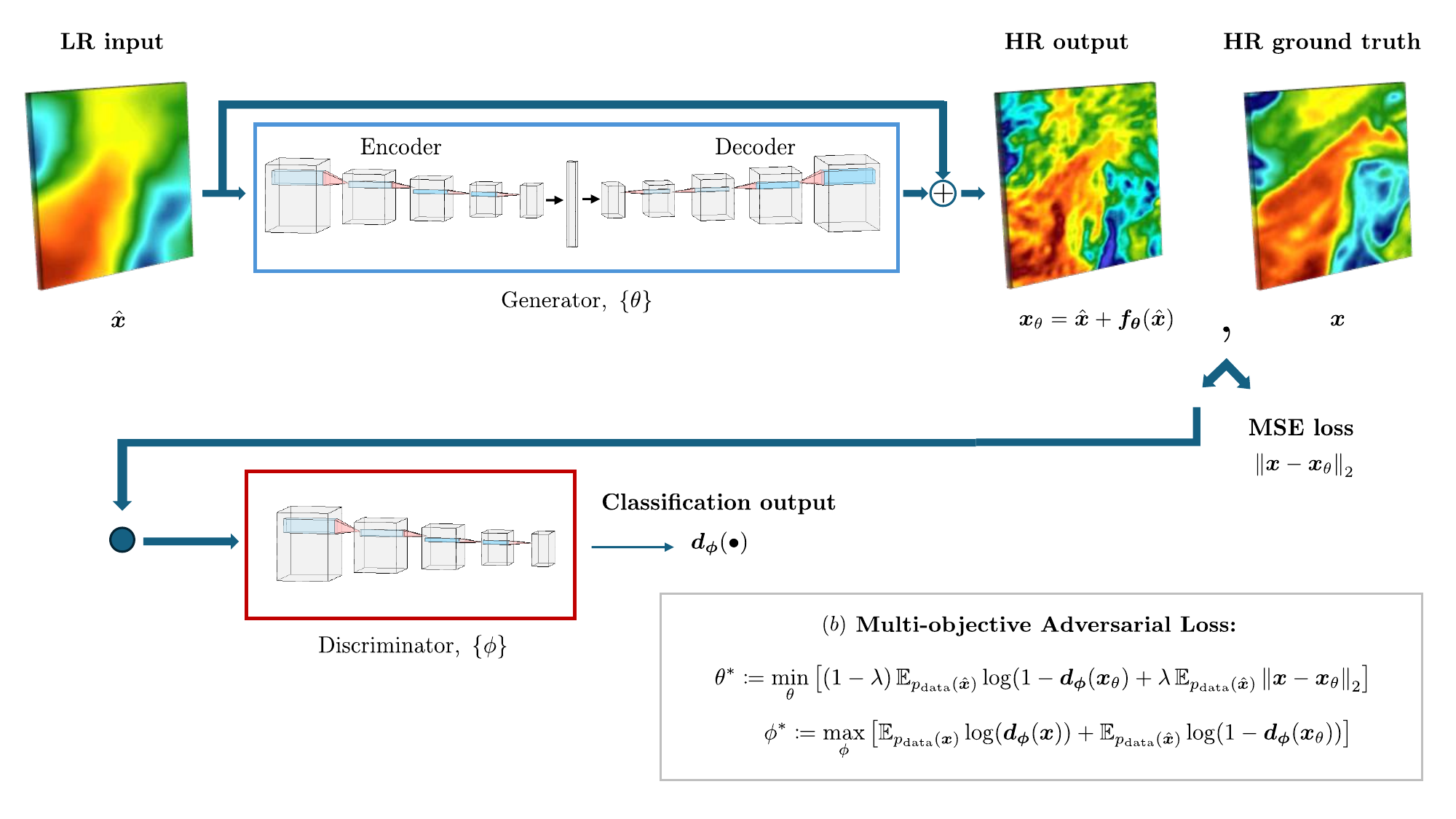}
\caption{The figure shows the model structure of the Conditional GAN (C-GAN) applied to the SST super-resolution problem.
The generator network, represented by the parameters $\theta$, consists of a residual network split into an encoder-decoder architecture to map the LR input map, $\hat{\bx}$, into a correction term that is added to the inputs themselves to obtain the HR reconstruction. The HR output, $\hat{\bx}+\bm{f}_\theta(\hat{\bx})$, is then compared to a ground truth data, $\bx$, both in terms of its pointwise MSE and its statistical comparison provided by the discriminator classification prediction, $\bm{d}_\phi(\bullet)$, where $\bullet$ indicates that the discriminator can take in input either the ground-truth data, $\bx$, or the generator reconstruction. The multi-objective adversarial loss functions used to identify the optimal parameters $\theta^*$ and $\phi^*$ of the two networks are defined as shown in panel (b). The Autoencoder (AE) network can be obtained from the same architecture of the Generator by optimizing on the loss function obtained with $\lambda=0$, hence without the discriminator's supervision.}\label{fig:figure2}
\end{figure*}

Here, the discriminator loss in eq.~\eqref{eq:discr_cond} remains unchanged, except that the expectation over the latent space distribution is now replaced with an expectation over the LR dataset, $ p_{\text{data}}(\hat{\bx}) $.  
The $ \lambda $ factor in the generator loss, eq.~\eqref{eq:gener_cond}, balances the trade-off between adversarial learning and MSE minimization, ensuring both statistical accuracy and grid-wise reconstruction quality. Specifically, when $\lambda = 0$, optimization is purely adversarial, focusing only on the statistical properties of the generated samples. When $\lambda = 1$, the generator is trained solely to minimize the $ L_2 $ norm between the reconstructed HR field, $ \bx_\theta $, and the corresponding ground-truth HR data, $ \bx $. 
The $L_2$ term is required to enforce conditioning on the specific ground-truth realization. 
For $\lambda \in (0,1)$, the loss balances statistical consistency and deterministic reconstruction, allowing for a trade-off between realistic sample generation and direct fidelity to observed HR data.  
Additionally, the adversarial term in the generator loss differs from the standard formulation, as it excludes the expectation over ground-truth data. This modification is made because the term depends only implicitly on the generator’s parameters and was also found to destabilize training when included in the generator loss \cite{pathak2016context, buzzicotti2021reconstruction}.  
Figure~\ref{fig:figure2} provides a detailed illustration of the network architectures implemented for both the generator and the discriminator models. The discriminator is a deep convolutional network to solve the binary classification problem, while the generator is implemented with a deep convolutional residual network whose output is added to the input as a correction term.
To analyze the impact of the statistical and conditioning terms in the generator loss, we compare the reconstruction quality of super-resolved fields obtained from two different training setups of the same generative model. We consider: (i) Autoencoder model (AE), trained with $ \lambda = 1 $, where the generator focuses solely on reconstruction loss without any adversarial/statistical optimization. (ii) Conditional GAN (C-GAN), trained with $ \lambda = 0.02 $, a value optimized to balance both statistical accuracy and reconstruction fidelity. The parameter $\lambda=0.02$ was determined empirically as providing an effective balance between statistical consistency and point-wise reconstruction accuracy; further refinements may yield marginal improvements depending on the specific application. This comparative analysis highlights the role of adversarial training in enhancing super-resolution performance while preserving the conditioning LR properties of the input fields.  

\section{Results}
\label{sec:results}

For clarity in the following analysis, we define the notation for the different SST fields as follows: $ \hr $: Ground-truth high-resolution tiles. $ \lr $: Input low-resolution tiles. $ \aen $: High-resolution reconstruction from the Autoencoder model. $ \gan $: High-resolution reconstruction from the C-GAN model.


\subsection{Single Tile Super-Resolution}
\label{subsec:tiles}

\begin{figure*}[htpb]
\begin{center}
\includegraphics[width=0.8\textwidth]{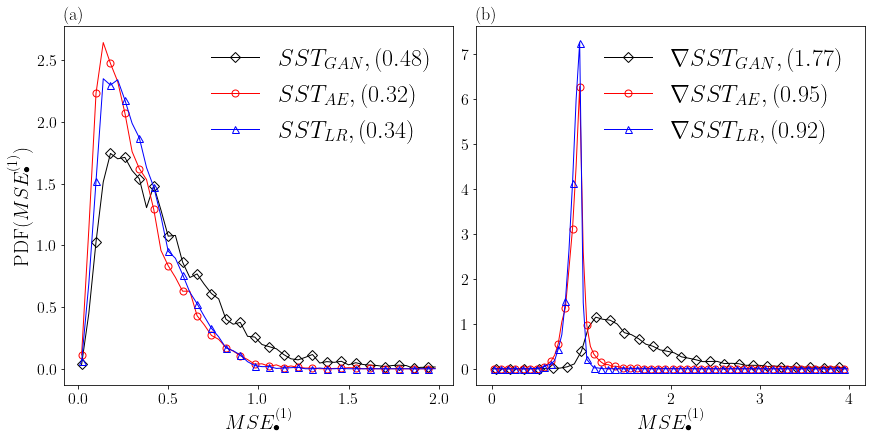}
\includegraphics[width=0.8\textwidth]{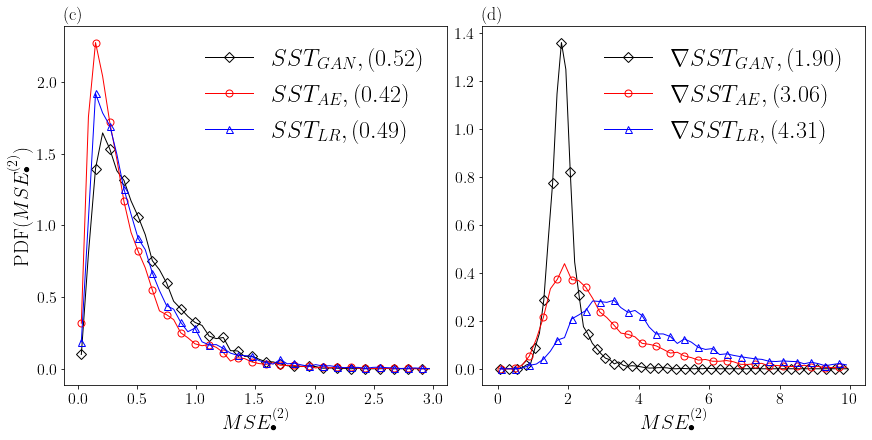}
\end{center}
\caption{Probability distribution functions (PDFs) of tile-wise normalized mean squared error, $MSE^{(1)}_\bullet$ and $MSE^{(2)}_\bullet$ panels (a) and (c) computed on the SST field for each tile of the test data, comparing the low-resolution input ($SST_{LR}$, blue), Autoencoder reconstruction ($SST_{AE}$, red), and C-GAN reconstruction ($SST_{GAN}$, black).  
Panels (b) and (d) are the PDFs of the same two normalized mean squared errors, but computed for the SST gradient magnitude ($\nabla SST_{\bullet}$). The numbers between the brackets in the legend are the averages of the corresponding PDFs.}\label{fig:figure3}
\end{figure*}

\begin{figure*}
  \centering
  \subfloat[a][Sample representative for the low MSE reconstructions.]{\includegraphics[width=1.\textwidth]{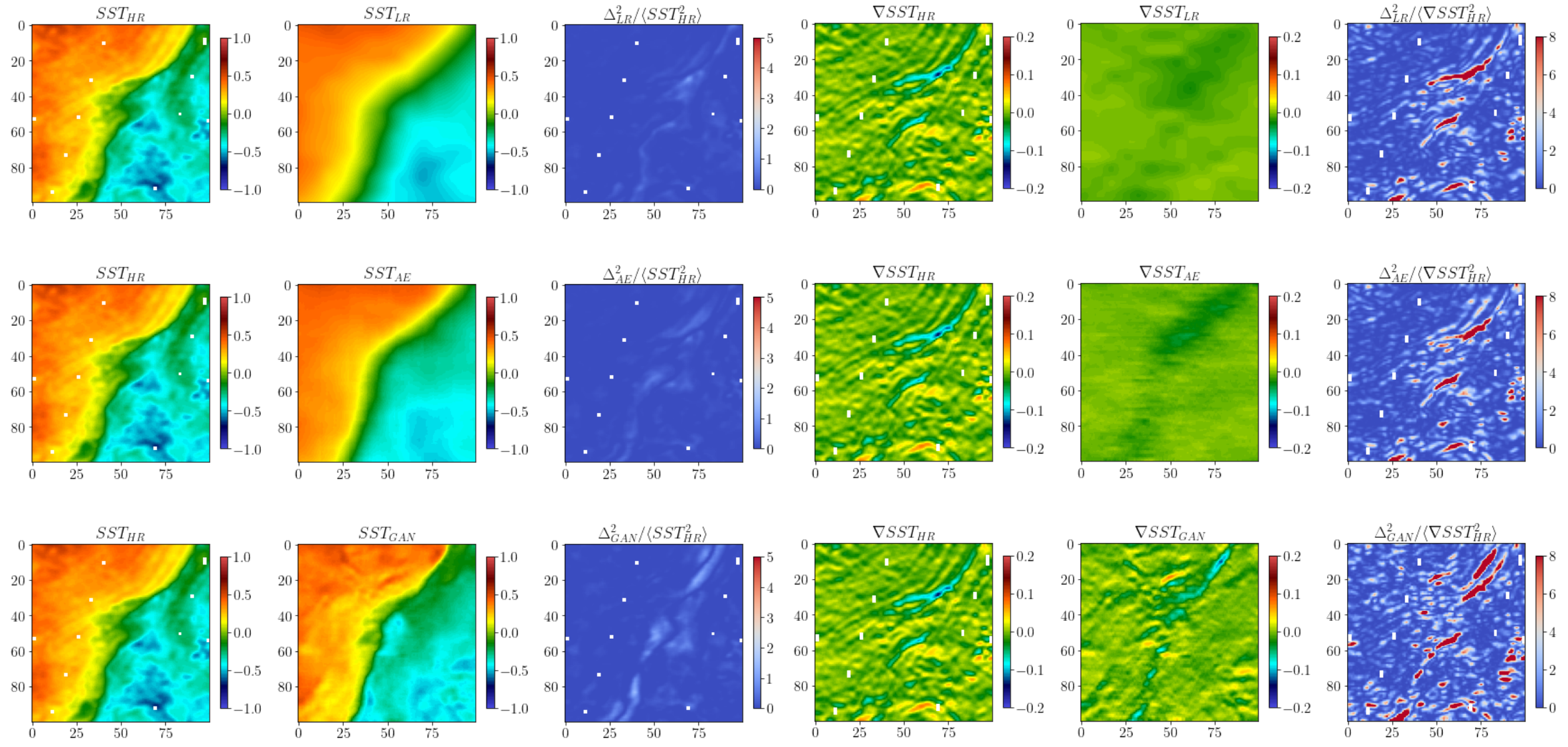}} \\
  \subfloat[b][Sample representative for the high MSE reconstructions.]{\includegraphics[width=1.\textwidth]{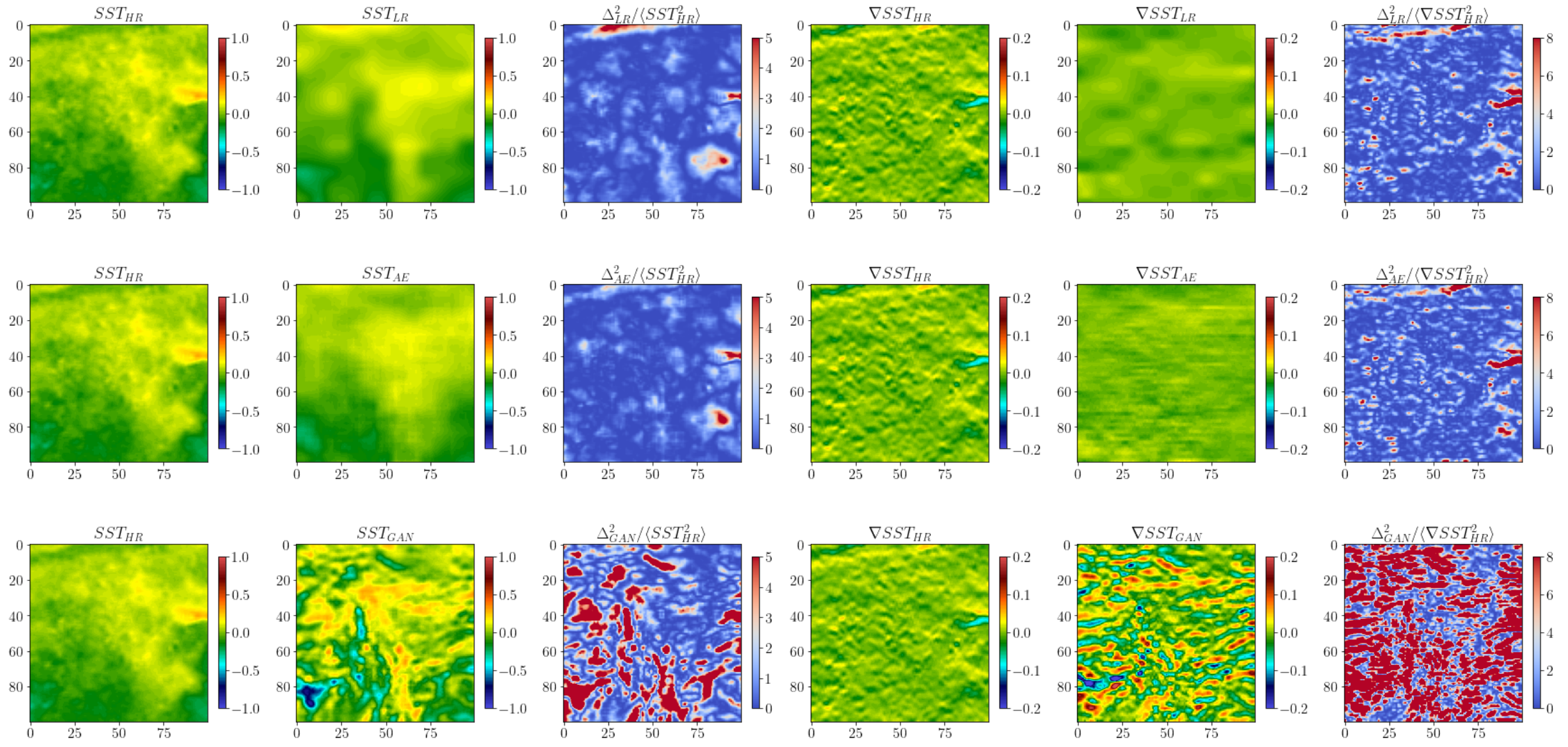}}
  \caption{Visual comparison of two representative tile reconstructions. Columns correspond to: the HR ground truth (first column), the model output (second column), MSE (third column), gradient field (fourth-fifth columns), and MSE for the SST gradients (sixth column).  
  In each of the two panels, there are three rows, with the LR input data, the the AE, and the C-GAN reconstruction respectively.} 
  \label{fig:figure4}
\end{figure*}

Figure~\ref{fig:figure3} presents the first quantitative evaluation, comparing the normalized MSE averaged over all tiles in the test dataset. This metric quantifies the improvement in point-wise reconstruction accuracy by measuring the difference between the HR ground truth and the different reconstructions. In this work we considered two different normalizations, (1) a normalization with respect to the mean value of the squared ground truth data and (2) a normalization which takes into account both the ground truth and the reconstructed data:
\begin{align}
    \text{MSE}^{(1)}_\bullet &= \frac{\langle (SST_\bullet - \hr)^2 \rangle}{\langle \hr^2 \rangle}, \\ \text{MSE}^{(2)}_\bullet &= \frac{\langle (SST_\bullet - \hr)^2 \rangle}{\langle SST_\bullet^2 \rangle^{1/2} \langle \hr^2 \rangle^{1/2}}
    \label{eq:MSE}
\end{align}
where with $SST_\bullet$ we indicate one of the three possible reconstructions given by the OI low-resolution maps or the HR reconstructed maps from AE or the C-GAN models. The average operation $ \langle \cdot \rangle = \frac{1}{A} \int_{A} (\cdot) dx_{lat}dx_{lon} $ is the spatial average computed on the area $A=100km^2$ of each tile from the test dataset.\\ \\
\noindent
Figure~\ref{fig:figure3}(a) and (c) show the probability density functions (PDFs) of the tile-wise errors $MSE^{(1)}$ and $MSE^{(2)}$, respectively, computed on the SST fields for the low-resolution input ($SST_{LR}$), the Autoencoder reconstruction ($SST_{AE}$), and the C-GAN reconstruction ($SST_{GAN}$). The bracketed values reported in the legends indicate the mean of each corresponding distribution. 
In panels (a) and (c) the errors for LR input and the AE reconstruction exhibit a very close distribution, with the AE showing a slight reduction in the mean error, suggesting a consistent, albeit modest, improvement over the low-resolution input. The C-GAN, on the other hand, yields a broader error distribution, indicative of its multi-objective optimization. Namely, the statistical optimization comes at the expense of higher point-wise errors, about $16\%$ higher than the AE in panel (a).
These differences are less pronounced in the case of $MSE^{(2)}$, panel (c), where the three methods yield closer distributions. Yet, the AE reconstruction maintains a slight improvement in its average performance, consistently achieving the lowest mean error.

Panels~(b) and (d) present the corresponding PDFs for the SST gradients fields ($\nabla SST_\bullet$) using the same error normalizations. For $MSE^{(1)}$ shown in panel (b), the trends are consistent with those seen for the SST fields. The AE and LR inputs have similar distributions, while the C-GAN again produces a broader spread. Interestingly, the mean value for the C-GAN in this panel approaches 2, which represents the asymptotic value one would expect for a reconstruction that statistically matches the ground truth but is uncorrelated point-wise. This behavior reflects the fact that the C-GAN introduces localized fluctuations to restore statistical fidelity, even if they are not phase-aligned with the ground truth data. It is important to notice that in this case, panel (b), the lower mean values observed for the AE and LR cases do not reflect an improvement in their gradient reconstruction accuracy, but rather the smoothness of the predicted fields. In fact, a perfectly smooth field with zero gradients would yield $MSE^{(1)} = 1$ by construction. 
This interpretation is further supported by the results in panel (d), where $MSE^{(2)}$ is computed for the same gradient magnitudes. In this setting, the C-GAN, which gets the same mean error around 2, now achieves the best performance among the three methods, highlighting its superior ability to recover small-scale in a statistical sense. In contrast, the underestimation of the gradients in both the LR input and AE reconstructions results in increased error values, as the denominator of $MSE^{(2)}$ is now sensitive to the artificially low SST gradient magnitude introduced into the reconstructed data.

To complement the PDF-based error analysis, Figure~\ref{fig:figure4} provides a visual comparison of two representative tiles: one selected from the low-MSE (panel a), and another from a high-MSE (panel b) sample of the C-GAN reconstruction. For each example, the first row compares the HR ground truth with the LR input tiles; the second row compares the HR and AE reconstructions; and the third row compares the HR and C-GAN results. In the low-error case, a dominant large-scale structure is present in the LR input, effectively constraining both model outputs. Indeed, even in the C-GAN case, the normalized tile-wise MSE remains low (on the order of 0.2), and the gradient fields also exhibit spatial coherence with the HR reference. In contrast, the high-error example (panel b) lacks strong, large-scale features in the LR input. This results in less significant conditioning and gives the C-GAN greater freedom to introduce small-scale stochastic variability. This results in a significantly higher MSE (around 2), with pronounced pointwise discrepancies in the gradient field, where C-GAN-generated fluctuations are visibly out of phase with the HR ground truth. In this case, the AE reconstruction has an advantage over the C-GAN by avoiding the introduction of small-scale fluctuations.\\ 
\\ \noindent
To further compare the two models, we analyze their reconstructions and errors across scales by performing a spectral decomposition in Fourier space.
\begin{figure*}[htpb]
\includegraphics[width=1\textwidth]{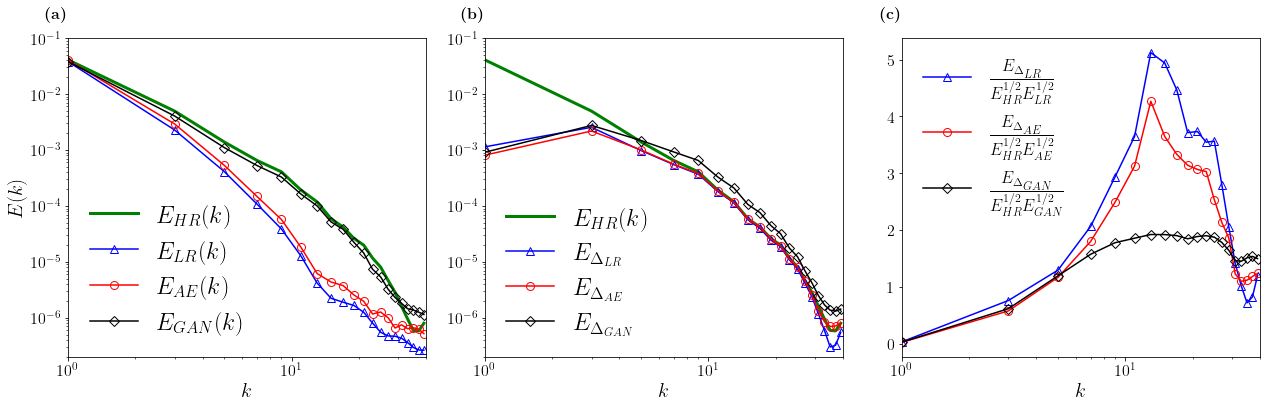}
\caption{Spectral analysis of SST reconstructions. Panel (a) Power spectral density $E(k)$ of the high-resolution SST field ($E_{HR}$, green solid line) compared with the low-resolution input ($E_{LR}$, blue triangles), the AE reconstruction ($E_{AE}$, red circles), and the C-GAN reconstruction ($E_{GAN}$, black diamonds). Panel (b) Power spectra of the reconstruction errors, $E_{\Delta}(k)$ for all fields.  
Panel (c) Normalized error spectra, showing the relative deviation of each reconstruction method from the HR reference. \label{fig:figure5}}
\end{figure*}

\begin{figure}[htpb]
\begin{center}
\includegraphics[width=0.5\textwidth]{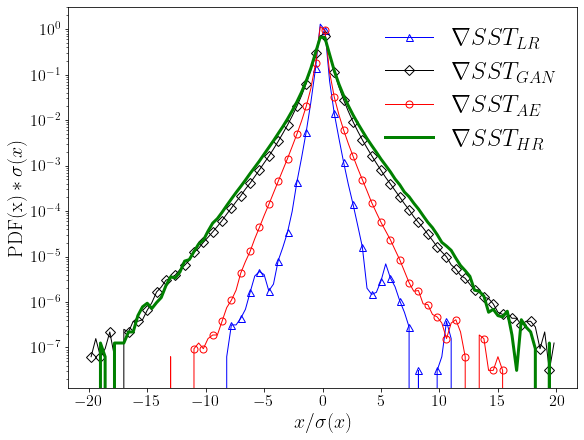}
\end{center}
\caption{Standardized PDF of SST gradients.  
The gradients of the HR reference field $\nabla SST_{HR}$ are reported in green solid line, LR gradients ($\nabla SST_{LR}$), are in blue triangles, the AE reconstruction gradients ($\nabla SST_{AE}$) are in red circles, and the C-GAN gradients ($\nabla SST_{GAN}$) are in black diamonds.\label{fig:figure6}}
\end{figure}

Figure~\ref{fig:figure5} presents the spectral decomposition of the $\hr$, $\lr$, and the reconstructed field, each compared with the spectra of the difference measured with respect to the ground truth HR data, namely $\Delta_\bullet=SST_\bullet - \hr$, where the $\Delta_\bullet$ represents as always one of the three cases, LR, AE and C-GAN.\\
Since we are now analyzing the spectral properties of the fields, two precautions are taken:  
(i) We select only gap-free tiles (with $100\%$ data availability), counting approximately $400$ tiles in the test dataset. (ii) A Hanning window is applied to enforce periodicity in latitude and longitude directions. Both these points avoid the presence of spurious Gibbs artifacts in the spectral analysis.   
The 1d Hanning taper function used in this work is defined as,

\begin{equation*}
    h(x) =
\begin{cases} 
\frac{1}{2} \left(1 - \cos\left(\frac{\pi \, x}{n_{taper} - 1}\right) \right), & 0 \leq x < n_{taper} \\[10pt]
1, & n_{taper} \leq x < n_{grid} - n_{taper} \\[10pt]
\frac{1}{2} \left(1 - \cos\left(\frac{\pi (n_{grid} - x - 1)}{n_{taper} - 1}\right) \right), & n_{grid} - n_{taper} \leq x < n_{grid}
\end{cases}
\end{equation*}

where $x$ is the index of the taper function, $n_{grid}$ is the total length of each tile (in our case $n_{grid}=100$), $n_{taper}$ is the tapering width, here set to $10$. As defined, the taper function smoothly transitions from $0$ to $1$ at the start and transitions back to $0$ at the end in $n_{taper}$ grid points, while leaving untouched the central region of the tile. The 2d taper function in latitude and longitude is simply obtained as the outer product of two 1d functions, $h(x_{lat},x_{lon}) = h(x_{lat}) \otimes h(x_{lon})$. Periodicity is then obtained for each of the fields by a point-wise multiplication with the 2d Hanning taper function, i.e. $\widetilde{SST}_{\bullet}(x_{lat},x_{lon}) = SST_\bullet(x_{lat},x_{lon}) h(x_{lat},x_{lon})$.
The Fourier spectrum is then computed as, 
\begin{equation}
    E_\bullet(k) = \int_{|\bk| = k} \left| \widehat{SST}_\bullet(\bk) \right|^2 d\bk
    \label{eq:spectrum}
\end{equation}
where $\widehat{SST}_\bullet(\bk) = \text{FT}[\widetilde{SST}_\bullet(\bx)]$ is the Fourier transform of the corresponding real space periodically smoothed SST field.
Figure~\ref{fig:figure5}(a) presents the spectra of the ground-truth $\hr $ (solid line green color), the reconstructed fields, $ \lr $ (blue triangles), $ \aen$ (red circles), and $\gan$ (black diamonds).  
Among the reconstructions, only $ \gan $ maintains the same energy content as the ground truth across the full range of resolved scales. In contrast, both $ \lr $ and $ \aen $ exhibit significantly lower energy, by nearly an order of magnitude, across all scales except at the lowest dominant wavenumbers, $|\bk| \le 3$.  
In Figure~\ref{fig:figure5}(b), we examine the Fourier spectra of the error fields. It becomes evident that having a more energetic reconstruction at high wavenumbers, the C-GAN generates larger errors at small-scales, whereas for the input smooth data $ \lr $ as well as for the $ \aen $ reconstructions, the spectra of the error at high-wavenumbers equals $E_{HR}(k)$, since the reconstructed dynamics is completely negligible.  
Panel (c) of the same figure highlights this effect by showing the normalized spectra of the errors, defined as $E_{\Delta_\bullet}/(E_{\bullet}^{1/2}E_{HR}^{1/2})$. 
The relative deviation of each reconstruction method from the HR reference shows that the C-GAN reconstruction is the only one improving at high-wavenumbers in the spectrum reconstruction. 
A closer inspection of the normalized spectra reveals that the improvements in $ \aen $ over $ \lr $ come primarily from an optimization of the first three wavenumbers, rather than from a real super-resolution enhancement.\\ \\
\noindent
Before analyzing the SST reconstruction over the entire Mediterranean region, we first perform a statistical comparison on individual tiles. To focus on small-scale fluctuation, figure~\ref{fig:figure6} presents the probability distribution functions (PDFs) of SST spatial gradients computed across the entire test dataset. The results show that only the C-GAN reconstruction can produce SST gradient magnitudes that are comparable to those found in the ground truth distribution.  

\begin{figure*}[htpb]
\includegraphics[width=1.\textwidth]{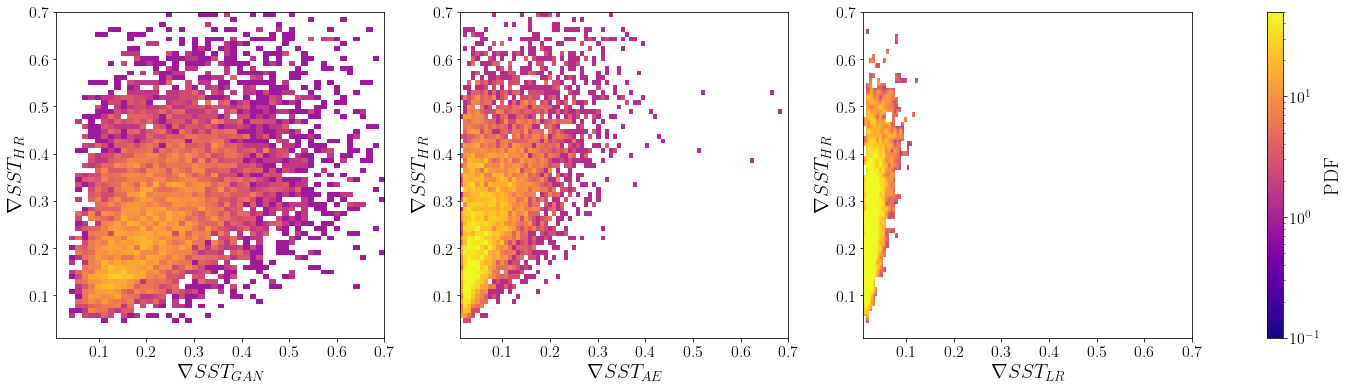}
\caption{Scatterplot of the maximum SST gradient measured in each tile in the HR ground truth fields ($\nabla SST_{HR}$, vertical axis) versus the corresponding reconstructions: (left panel) C-GAN ($\nabla SST_{GAN}$), (middle panel) AE ($\nabla SST_{AE}$), (right panel) LR ($\nabla SST_{LR}$). Colors represent the PDF of the maximum SST gradients on a logarithmic scale. 
\label{fig:figure7}}
\end{figure*}

The AE reconstruction only slightly improves the LR input statistics, while significant discrepancies with the ground truth HR distribution remain. This quantitatively demonstrates the limitation of the AE approach in reproducing the rare but extreme fluctuations far from the standard deviation.\\ \\ 
\noindent
In figure~\ref{fig:figure7}(a-c), we present the scatter plots of the maximum SST gradients predicted by the different reconstructions compared to the HR ground truth data ($\nabla \hr $). This analysis quantifies the ability of different models to accurately predict the probability of observing small-scale extreme events, conditioned on the LR SST configurations. This provides an estimate of the conditional probability modeled by the different approaches in terms of the observed maximum SST gradient.

The right panel shows $\nabla \gan$ vs. $\nabla \hr$, the middle panel shows the AE results, and the right panel shows the scatterplot obtained between the LR and the ground truth HR data.
The LR data strongly underestimate the gradients measured in the HR observations, with most of the data points concentrated near the vertical axis. 
Again, we see that the C-GAN super-resolution provides the most accurate approximation of the true maximum SST gradients. In this case, the data distribution is more spread and better aligned along the scatterplot diagonal, showing that C-GAN is effective in recovering the range of the gradient distribution of the high-resolution observations.
The AE reconstruction improves on LR by capturing a wider range of gradient values. However, there remains a noticeable difference with respect to the HR ground truth, as the AE produces a systematic underestimation of the SST gradients. 


\subsection{Mediterranean Map reconstruction}
\label{subsec:med}

\begin{figure*}[htpb]
\includegraphics[width=\textwidth]{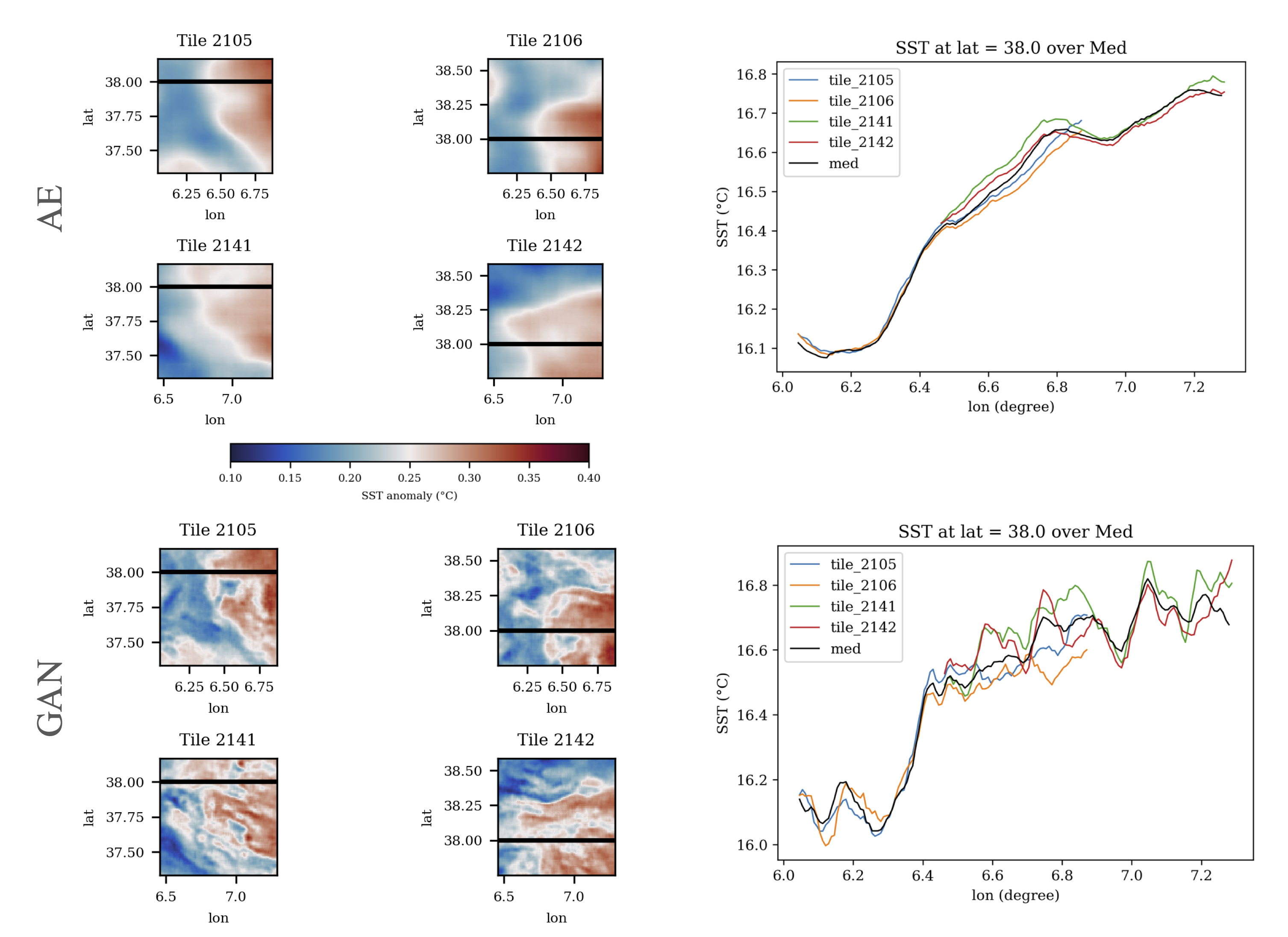}
\caption{On the left, four adjacent tiles at the common latitude $38^\circ$N obtained from the super-resolved AE (top) and GAN (bottom) outputs obtained from the LR SST fields. On the right, the comparison of the derived SST profiles for the single tiles with the final reconstructed SST distribution over the transect selected. \label{fig:comparison_reconstruction}}
\end{figure*}

We now extend the SST reconstruction analysis beyond individual tiles by moving to the regional maps of the Mediterranean Sea, obtained by combining and gluing different super-resolved tiles. The merging process consists of calculating a linear combination of all the overlapped pixels weighted according to their distance from the center of the tile. The more central a pixel is, the more importance it will have in the reconstruction. In fact, each tile is multiplied by a weighting matrix representing a smooth, symmetrical 2D pyramid-like function centered in the middle of the matrix. The weights increase gradually from the edges (starting from 0) toward the center, reaching a maximum equal to 0.25 at the midpoint. Each pixel is then the linear combination of the four values obtained within the tiles containing that pixel. Then, the local spatial mean initially removed is added back to recover the final SST values over the Mediterranean Sea. All these operations ensure smooth transitions between adjacent reconstructions, as illustrated in Fig. \ref{fig:comparison_reconstruction} for both AE and GAN outputs.
 
The eight visualizations on the left of the image (four for each model) show individual super-resolved tiles at different adjacent locations, highlighting an overlapping transect with a black horizontal line at the common latitude $38^\circ$N. On the right, the corresponding SST profiles (in black) over the Mediterranean Sea at the selected latitude are shown, alongside the four SST distributions obtained from the tiles. The reconstructed SST shows how the merging procedure used minimizes discontinuities and artifacts, ensuring a smooth and consistent reconstruction of the SST field. This is crucial to properly compare the output of the networks with the ground-truth data over the whole basin.

\begin{figure*}[htpb]
\includegraphics[width=0.9\textwidth]{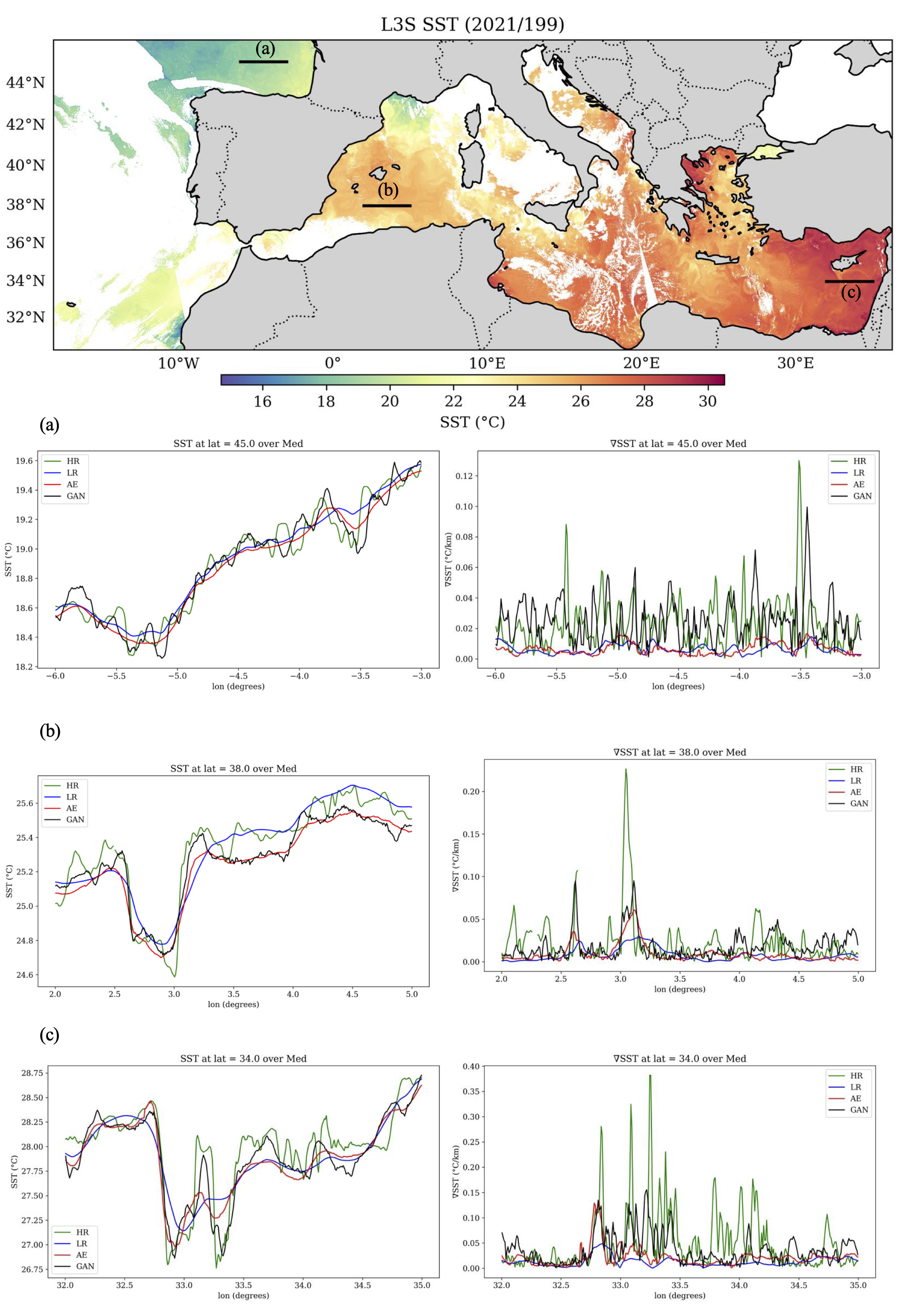}
\caption{The top panel shows the high-resolution L3S SST data 18/07/2021 (julian day = 199). Black lines over the image highlight the three transects over which we present the SST variations (left column) and SST gradients (right columns) for the HR, LR, AE and GAN reconstructions in function of the longitudes in panels (a), (b) and (c). \label{fig:transects}}
\end{figure*}   

\begin{figure*}[htpb]
\includegraphics[width=\textwidth]{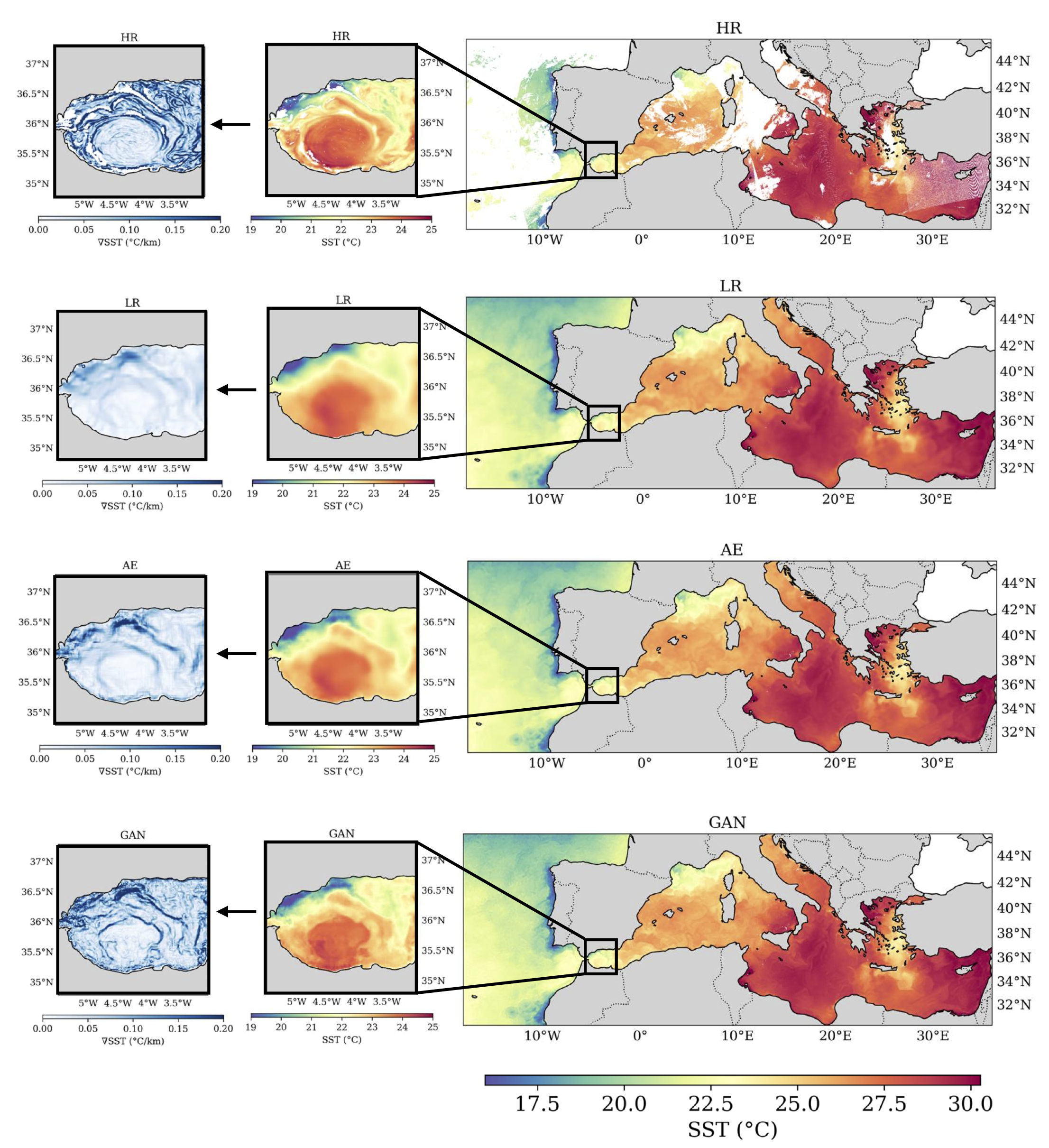}
\caption{SST distribution over the Mediterranean Sea on the 5 August 2021 obtained from the high-resolution ground-truth data (HR), the optimal interpolated product (LR), the autoencoder (AE) and the generative adversarial network (GAN) outputs with a focus on a structure-rich area in the Albor\'an Sea (delimited by the coordinates 34.8–37.3$^\circ$N, 5.5–3$^\circ$W) with the correspondent SST spatial gradients. \label{fig:jday217_FINAL}}
\end{figure*}

While it is well known how the optimal interpolation of the SST data tends to severely smooth the fields even when direct observation is present \cite{nardelli2013high, fanelli2024deep}, we want to visualize how the different networks handle the reconstruction of small-scale variations. Figure~\ref{fig:transects} shows the SST profiles and gradients as a function of longitude over three transects at different fixed latitudes ($34, 38$ and $45^\circ $N) over the basin. The HR satellite observation (green), created following the same process applied to the target dataset of the networks, serves as the reference, while the LR data (blue) represents the input for the two data-driven models: the AE (red) and the C-GAN (black).  As mentioned above, the LR profile appears significantly smoother than the HR data, lacking fine-scale variability due to the averaging effects of the gap-filling process. The AE reconstruction closely follows the LR trend, introducing minor refinements but failing to recover small-scale details. In contrast, the C-GAN reconstruction shows enhanced fluctuations that capture well the same small-scale statistics as in the ground truth data, as can be seen from the SST gradients comparison on the right panels. Although, as discussed in the previous sections, a point-wise reconstruction of the small-scales is not possible by any method, since the problem is not well posed. As a consequence, the C-GAN reconstructed fluctuations only match the real statistics of the SST field, but not the exact profile. In general, AE provides a smooth reconstruction, lacking the high-frequency structures. \\
\noindent
To better visualize the different reconstructions of the two models analyzed in this work, Figure \ref{fig:jday217_FINAL} compares the SST distribution over the Mediterranean Sea on 5 August 2021 obtained from the high-resolution ground-truth data (HR), the optimal interpolated product (LR), the autoencoder (AE) and the generative adversarial network (GAN) outputs, with a focus on a structure-rich area in the Albor\'an Sea with the correspondent SST spatial gradients. 

\begin{figure*}[htpb]
\includegraphics[width=\textwidth]{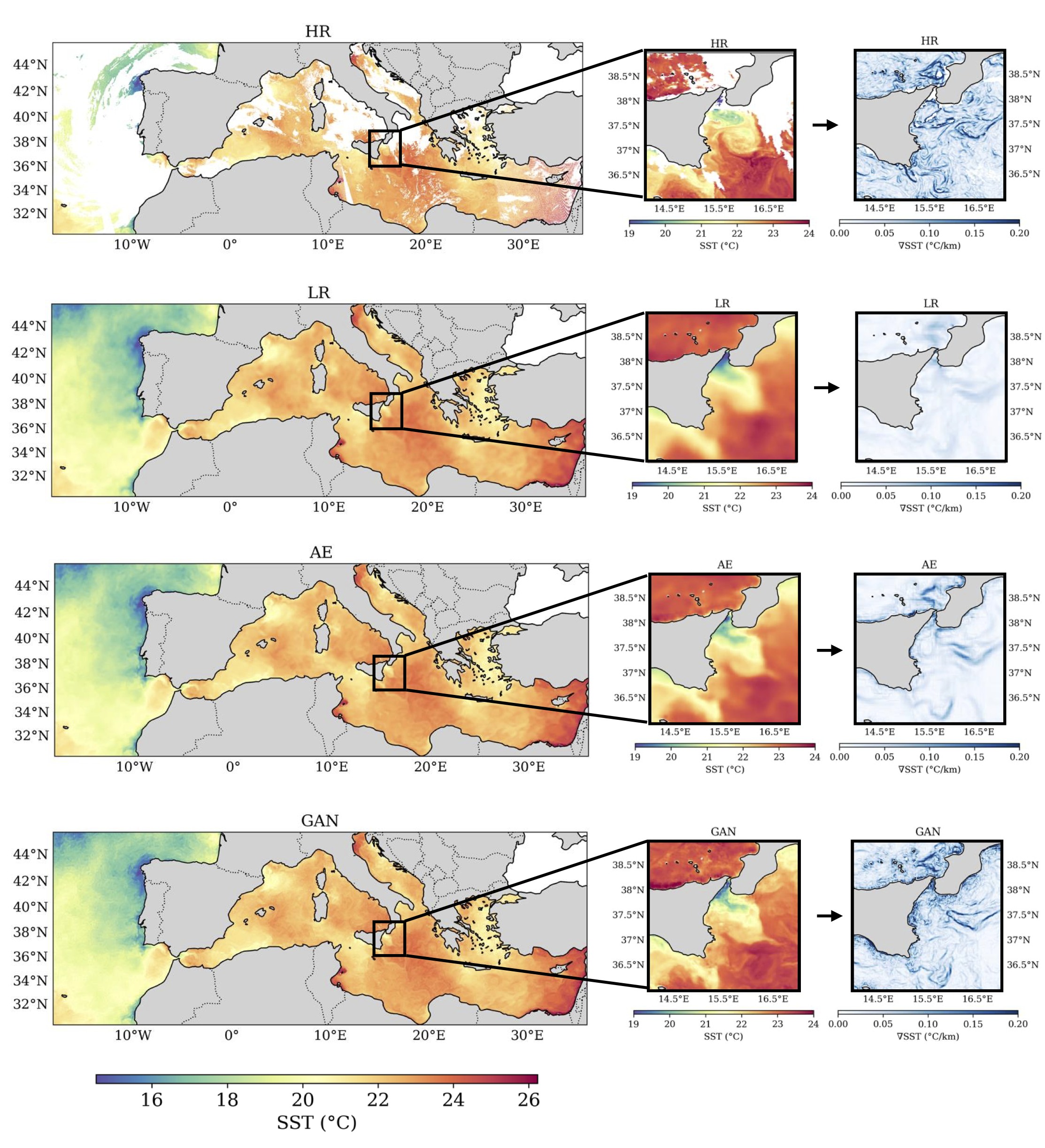}
\caption{SST distribution over the Mediterranean Sea on the 12 June 2021 obtained from the high-resolution ground-truth data (HR), the optimal interpolated product (LR), the autoencoder (AE) and the generative adversarial network (GAN) outputs with a focus on the sea adjacent the eastern Sicily and southern Calabria (delimited by the coordinates 36–39$^\circ$N, 14–17$^\circ$E) with the correspondent SST spatial gradients. \label{fig:jday163_FINAL}}
\end{figure*}

\noindent
The different scales reproduced by the four representations quickly become evident. The $1/100^\circ$ effective spatial resolution offered by the HR observations is lost in the LR image, which still captures the large-scale dynamics but loses most of the meso- and sub-mesoscale variability over the basin. Similar behavior can be observed in the AE SST field as opposed to the reconstruction obtained from the GAN output. The latter exhibits diffusely small-scale variability, particularly noticeable in the Atlantic Ocean. The differences can be understood by looking at the regional zoom on the left, which shows the SST gradients representation in the same area. The persistent circulation of the Western Albor\'an Gyre (WAG), delimited by the Moroccan coast on the south and the cold insertion of the Atlantic Jet (AJ) on the north is sharply visible in the HR image while it appears quite smooth in the LR optimal interpolated field. The AE network manages to recover the higher intensity of some fronts of the input image without adding any missing structure, effectively acting as a gradient enhancement. On the other hand, the GAN introduces new small-scale variability in addition to improving the high-frequency fluctuations. However, it reduces the WAG dimension and misplaces some of the surrounding small-scale structures.\\
\noindent
Figure~\ref{fig:jday163_FINAL} shows the four SST images over the Mediterranean Sea on 12 June 2021 with a zoom over the sea adjacent to eastern Sicily and southern Calabria (Italy). The general observations regarding the sharpness/smoothness of the different fields made for Fig.~\ref{fig:jday217_FINAL} still stand. However, the regional zoom shows a different situation. Looking at the ground-truth data, we can observe two separate eddies just below the Strait of Messina, reconstructed as a unique colder structure by both networks. Even if they both enhance the intensity of the fronts, they cannot recover the correct position of the mesoscale structures, while only the C-GAN architecture can improve the small-scale variability of the reconstructed field. 

\begin{figure}[htpb]
\begin{center}
\includegraphics[width=0.5\textwidth]{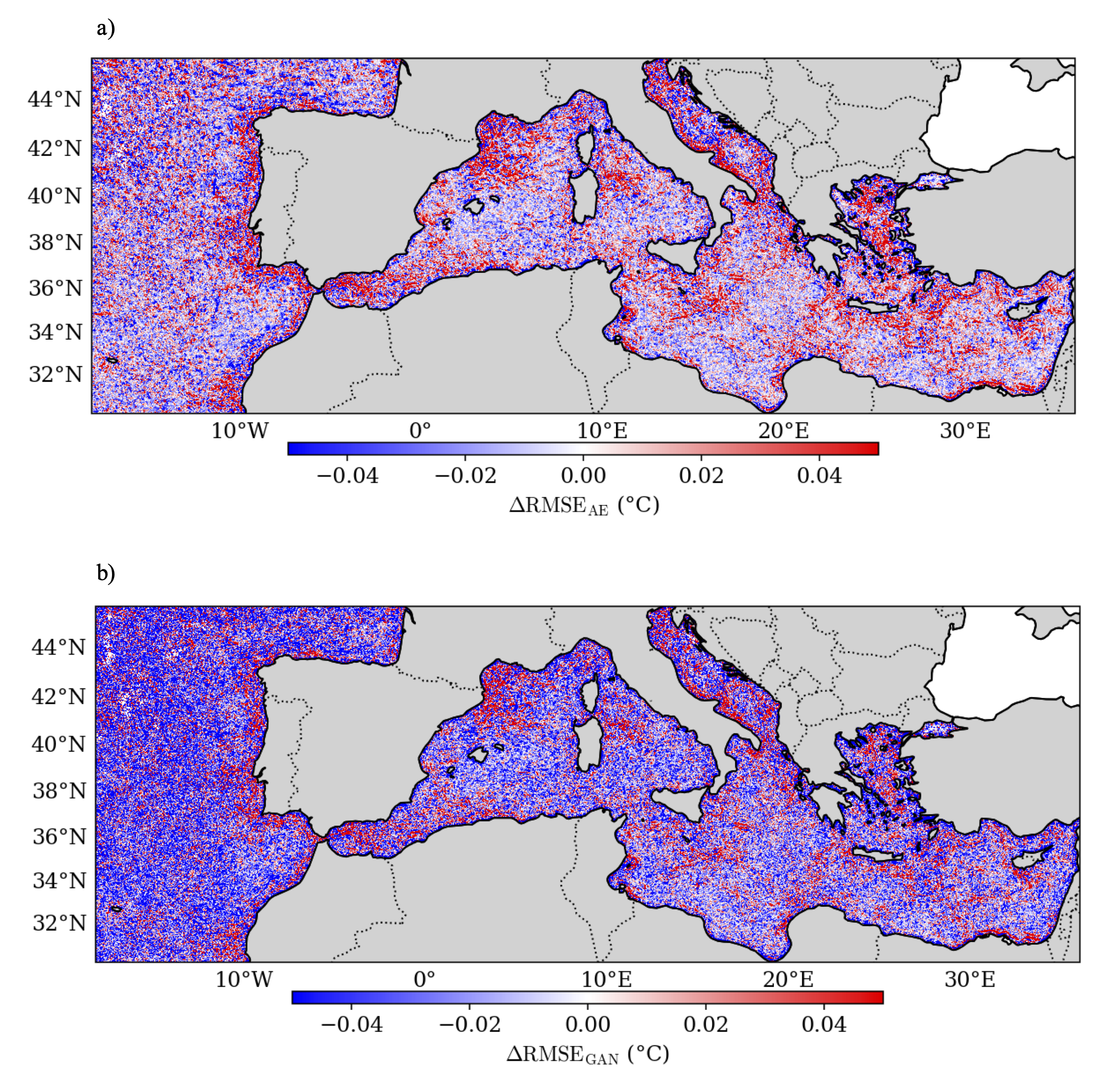}
\end{center}
\caption{Comparison of the performance of both (a) AE and (b) C-GAN SST reconstruction against the optimal interpolated fields with respect to the L3S data measured by Sentinel-3A and 3B satellites during 20 days of 2021. Positive red (negative blue) values show an improvement (deterioration) of the networks reconstruction with respect to the optimal interpolated data.
\label{fig:errors_med}}
\end{figure}

\noindent
From the examples presented seems that the neural networks used act like gradient enhancement, but they are not always able to capture the exact features of the ocean. This analysis is confirmed, as shown in Fig.~\ref{fig:errors_med}, by the results of the root mean square error difference between the input and the reconstructed SST for each pixel, defined as,
\begin{multline}\label{eq:DRMSE}
    \Delta \text{RMSE}_\bullet = \sqrt{\langle (\lr - \hr)^2 \rangle}_t - \\
    - \sqrt{\langle (SST_\bullet - \hr)^2 \rangle}_t,
\end{multline}
where the $\langle \cdot \rangle_t$ represents the temporal average.  

Positive (red) values indicate regions where the data-driven reconstruction reduces MSE compared to the LR input, while negative (blue) values indicate a deterioration in reconstruction accuracy for the AE (Fig.~\ref{fig:errors_med}a) and the C-GAN (Fig.~\ref{fig:errors_med}b) outputs. The results confirm the distinct behaviors for the two models:
\begin{enumerate}
    \item The AE reconstruction improves the MSE of the SST field at some points, enhancing the intensity of the smooth field produced by the OI but failing to capture the small scale variability.
    \item The C-GAN reconstruction, even recovering fine-scale features as demonstrated in Sec.~\ref{subsec:tiles}, increases MSE almost everywhere due to the wrong position/shape of the generated structures.
\end{enumerate}


\section{Conclusion}
In this work, we investigate the single-image super-resolution of low-resolution sea surface temperature (SST) fields derived from interpolated satellite observations. Owing to the under-determined nature of the problem, super-resolution is a deterministically ill-posed task and, depending on the final application of super-resolved data, it may be useful to frame it in statistical terms. 
\\ \noindent
We find that deterministic models trained solely with an $L_2$ loss, such as the Autoencoder (AE), improve large-scale agreement but fail to enhance the energetic content at small scales. As a result, they are unable to recover the statistics of subgrid variability. In contrast, generative models that incorporate statistical learning objectives can approximate the distribution of small-scale features conditioned on large-scale inputs. Specifically, we focus on the Conditional GAN (C-GAN), trained with a hybrid loss that combines an adversarial (statistical) term with $L_2$ conditioning. This architecture still ensures fidelity to the observed large-scale structures and restores small-scale variability in a statistically consistent manner, without attempting to reconstruct exact realizations. \\
\\ \noindent
Our results demonstrate that statistical super-resolution is achievable using modern generative methods. However, while not explicitly shown here, this approach may produce super-resolved features that are inconsistent over time, similar to the spatial inconsistencies observed in overlapping image tiles. As such, future efforts will be dedicated to the set-up of multiple-image super-resolution models, which will be able to guarantee a much higher temporal consistency, also eventually abandoning the tiling approach at inference phase. This represents a promising path forward for improving the physical realism of gap-filled satellite data, albeit at a significantly higher computational cost that will necessitate dedicated resources. Future work could further extend this approach using more advanced probabilistic frameworks, such as diffusion models, to further enhance the stability and expressiveness of learned multiscale geophysical dynamics.


\section*{Acknowledgments}
CF has been financially supported by the Italian national project ITINERIS (IR0000032 - ITINERIS, Italian Integrated Environmental Research Infrastructures System, D.D. n. 130/2022 - CUP B53C22002150006, Funded by EU - Next Generation EU PNRR - Mission 4 "Education and Research" - Component 2: "From research to business" - Investment 3.1: "Fund for the realisation ofan integrated system of research and innovation infrastructures".
This work has been partially supported by the Copernicus Marine Service through contract no. 21001L03-COP-TAC SST-2300 via the Provision of Sea Surface Temperature – Thematic Assembly Center (SST-TAC) Observation Products (SST-TAC) project and by the European Research Council (ERC) under the European Union’s Horizon 2020 research and innovation programme Smart-TURB (Grant Agreement No. 882340).

\bibliographystyle{unsrt}
\bibliography{references}

\vfill

\end{document}